\documentclass[%
reprint,
nofootinbib,
amsmath,amssymb,
aps,
pra,
noeprint,
nolongbibliography
]{revtex4-2}

\usepackage{graphicx} 
\usepackage[table,xcdraw]{xcolor} 

\usepackage{physics}
\usepackage{siunitx}
\AtBeginDocument{\RenewCommandCopy\qty\SI}

\usepackage{dsfont} 
\usepackage{mathtools}

{ 

\newcommand{\deceff}{\mathcal{D}}
\newcommand{\fidelity}{\mathcal{F}}
\newcommand{\seq}{\text{DD}}

\newcommand{\cxiii}{{}^{13}\text{C}}
\DeclareMathOperator{\inlineTr}{tr}

\DeclareSIUnit\gauss{G}

\newcommand{\idop}{\mathds{1}}
\newcommand{\defeq}{:=}
\newcommand{\eqdef}{=:}

\newcommand{\zx}{\mathcal{A}^x}
\newcommand{\zy}{\mathcal{A}^y}
\newcommand{\swap}{\text{SWAP}}
\newcommand{\iswap}{\text{iSWAP}}

\newcommand{\DeltaJ}{\Delta_{j}}
\newcommand{\TOop}{\mathcal{T}_\leftarrow}

\makeatletter
\newcommand{\fmarki}{*}
\newcommand{\fmarkii}{\ensuremath{\dagger}}

\renewcommand{\fmarki}{$\circ$}
\renewcommand{\fmarkii}{$\bullet$}

\def\@fnsymbol#1{{\ifcase#1\or \fmarki\or \fmarkii\or \fmarkiii\or \fmarkiv\or \fmarkv\or \fmarkvi\or \fmarkvii\or \fmarkviii\or \fmarkix \else\@ctrerr\fi}}
\makeatother

\begin{document}
	
\title{Towards quantum error correction with two-body gates for quantum registers based on nitrogen-vacancy centers in diamond}
\author{Daniel Dulog}
\email{daniel.dulog@uni-ulm.de}
\author{Martin B. Plenio}
\email{martin.plenio@uni-ulm.de}
\affiliation{Universität Ulm, Institut für theoretische Physik, 89081 Ulm}
\date{\today}
	
\begin{abstract}
	Color centers in diamond provide a possible hardware for quantum computation, where the most basic quantum information processing unit are nitrogen-vacancy (NV) centers, each in contact with adjacent carbon nuclear spins. With specifically tailored dynamical decoupling sequences, it is possible to execute selective, high-fidelity two-body gates between the electron spin of the NV center and a targeted nuclear spin. In this work, we present a method to determine the optimal execution time that balances the trade-off between fidelity and execution speed for gates generated by adaptive XY sequences. With these optimized gates, we use the nuclear spin environment as a code space for quantum error correction within a color center register.
\end{abstract}

\maketitle

\section{Introduction}
Since the emergence of the idea of quantum information processing and quantum computing, the physics community has been attempting the technically challenging experimental realization of these concepts on many different physical platforms, such as superconducting circuits, ion traps, photonic quantum computers, or solid-state spin-based architectures. The latter approach, in particular systems based on nitrogen-vacancy (NV) centers, provides a promising physical foundation for quantum computing due to their optical readout and manipulation capabilities as well as robust qubit realizations in terms of decoherence times \cite{doherty2013, childress2013}. 
Albeit architectures solely based on the electron spin of the NV center are conceivable, the adjacent $\cxiii$ nuclear spins can be used to support the electronic NV center qubit or even constitute the quantum register, while the electron spins act only as interaction mediators. Initial experiments with a carbon spin register have been conducted, utilizing a nearby spin ensemble to create small quantum error correction codes. Different methods were used to generate electron-nuclear two-spin operations, such as employing a microwave pulse sequence \cite{cramer2016, taminiau2014} (similar to the approach discussed here) or applying frequency-selective microwave pulses \cite{waldherr2014}.

 Achieving high-fidelity quantum operations in an electron-nuclear quantum register with multiple nuclei is inherently challenging. This difficulty arises because high fidelity for an operation on a subsystem primarily depends on spectral selectivity. Spectral selectivity, in turn, requires extended sequences in the time domain due to the time-bandwidth product limit. However, making a gate sequence too long has significant drawbacks. Firstly, excessively long gates slow down the overall computation speed of the device. Secondly, and more critically, prolonged sequences increase the exposure to relaxation and decoherence processes caused by imperfect shielding from the environment. These processes can degrade the fidelity that was achieved through the extended sequence needed for energy selectivity. Therefore, in designing any quantum gate, one must carefully balance the trade-off between the gate's fidelity and its execution time.
 
Unlike standard dynamical decoupling sequences like XY and CMPMG, introducing dynamical decoupling sequences with additional degrees of freedom in pulse positioning offers greater control over the shape and height of the filter function \cite{albrecht2015, albrecht2015a}. This increased control can enhance the fidelity of operations by improving selectivity and achieving more precise rotation angles.

 In this work we demonstrate time-optimized, high-fidelity operations between electron and nuclear spins using specifically engineered dynamical decoupling sequences called adaptive XY sequences (AXY) \cite{casanova2015, casanova2016}. These sequences achieve spectral selectivity by reducing the effective coupling strength and thus sharpening the filter function of the operation. Typically, the lower bound of the length of such a selective sequence is (roughly) chosen by an approximation condition determined by the validity of a rotating wave approximation (RWA). In this work we use a different protocol that makes use of the fact that non-RWA terms cause oscillations in the system’s evolution that act as a perturbation with respect to the target operation. It is crucial to note for what follows, that due to the periodic nature of the perturbation the error terms approach zero for specific discrete moments in time. A judicious choice of the gate time then may find a locally minimal infidelity even outside the parameter range suggested by the RWA condition. This method allows a precise selection of an optimal gate time and eliminates the need for extensive parameter scans and the cumbersome experimental gate tomography that these require.

Moreover, these two-qubit gates are used in a compact (2+1)-spin protocol in a structure derived from the well-known repetition code \cite{nielsen2010} which can protect the quantum register against one type of errors - in the case of this work a phase error. Through our demonstration, we showcase the efficacy of this protocol, constructed through a sequence of AXY-based gates, in reliably shielding or recovering single qubit quantum information against phase errors.

\section{High-fidelity two-qubit gates}
\label{sec:gates_basics}
\begin{figure*}
	\centering
	\includegraphics[width=.94\linewidth]{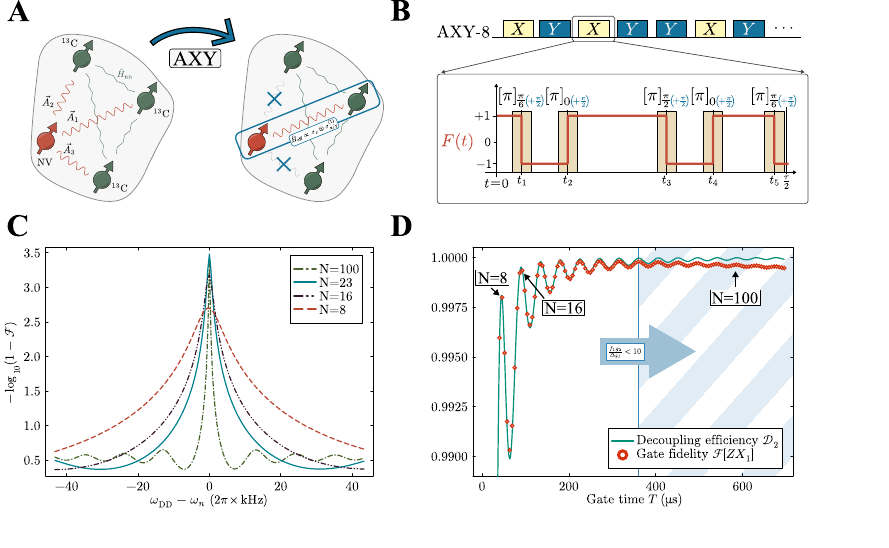}
	\caption{(a)  Effect of the AXY-based two-qubit gate on the quantum register. Both the targeted nuclear spin (here: spin 1) and the NV center undergo a two-qubit operation with the effective Hamiltonian $\smash{H_\text{eff} \propto \sigma_z \otimes I_\text{x/y}^{(1)}}$, while all other off-resonant hyperfine interactions are annihilated. (b) Illustration of the AXY-8 dynamical decoupling sequence. It consist of $X$ and $Y$ pulses just as the XY-8 sequence, but all $\pi$-pulses are replaced by composite pulses of length $\tau / 2$. The five sub-pulses are non-equidistant and allow specific tailoring of the modulation function $F(t)$. $Y$-pulses differ from $X$-pulses by a phase shift of $\pi/ 2$. (c) Logarithmic infidelity curves for the gate $\zx_1 (\pi / 2)$ scanned across the dynamical decoupling frequency $\omega_\seq$ for various sequence repetitions $N$. (d) Relationship between fidelity $\fidelity \left[\zx_1(\pi /2)\right]$ and decoupling efficiency $\deceff_2(T)$ for a three-spin system. The hatched area represents the parameter regime constrained by the coupling limit approximation \eqref{eq:approxCond1}.}
	\label{fig:fig1}
\end{figure*}
In this part we explain the basic principles of generating two-qubit gates with adaptive dynamical decoupling sequences, such as the AXY sequence, that are used throughout the work.

A color-center-based quantum processor may be comprised of registers that are physically realized by negatively charged nitrogen-vacancy (NV) centers in diamond; each surrounded by multiple 13C nuclear spins (Fig.~\ref{fig:fig1}A) which appear in natural abundance of approximately one percent. In such a register, the spins are coupled by dipolar hyperfine interaction, but the relatively weak internuclear interactions are disregarded as information transfer channels and will only induce crosstalk among the physical qubits on longer timescales. With the remaining electron-nuclear dipole-dipole interactions the system attains a star-shaped geometry and hence an all-to-one connectivity.

In this work, we propose an effective two-qubit operation between the electron and nuclear spin segments in a diamond-based quantum register by adaptive dynamical decoupling \cite{casanova2015, casanova2016}. An adaptive sequence, i.e. a sequence with the ability to continuously adjust the effective coupling strength, enables the generation of high-fidelity gates for several reasons. First, by decreasing or increasing the effective coupling strength, the sequence allows the shaping of its filter function (see Appendix \ref{sec:filter_functions}) and is therefore able to perform the fastest possible gate for a given spectral separation of the spin ensemble components. Second, for a two-qubit interaction $A \otimes B$ it allows to generate all evolution operators $e^{-i\varphi A \otimes B}$ for all values of the continuous variable $\varphi$, even though pulse sequences can only be evaluated a discrete number of times, i.e. ~multiples of their period. Finally, dynamical decoupling sequences provide protection against drive control errors or other extrinsic noise sources. The adaptive XY sequence is a modification of the XY sequence where each X and Y pulse consists of a Knill composite pulse sequence with varying time intervals between the $\pi$-pulses (see Fig.~\ref{fig:fig1}B) \cite{lidar2013, souza2011, ryan2010}. These variable times $\{t_i \}_{i=1}^{10}$ provide degrees of freedom for the tuning of the effective coupling strength.

A static external magnetic field is aligned with the NV center's axis $\hat{z}$, i.e.~$\vec{B} = B \hat{z}$, and furthermore time-dependent oscillating MW-fields are used for electron spin control. In secular approximation, the Hamiltonian of such a system reads 
\begin{align}
	\label{eq:general_sys_hamiltonian}
	H = ~&D S_z^2 - \gamma_e B S_z \nonumber\\ &- \sum_j \gamma_j B I_j^z + S_z \sum_j \vec{A}_j \cdot \vec{I}_j  + H_c + H_\text{nn}.
\end{align}
Here, $H_\text{nn}$ is the dipolar internuclear coupling between the $\cxiii$-spins and $H_c$ is the microwave control field. The internuclear coupling $H_\text{nn}$ is included in the simulations but will be discarded in the analytical discussion. Assuming a large enough splitting between the electron spin $\ket{\pm1}$-states compared to the control field amplitude, it is convenient to treat the NV center as an effective two-level system by transforming $S_z \rightarrow \frac{m_s}{2} (\sigma_z + \idop)$ with $\sigma_z = \dyad*{m_s}{m_s} - \dyad*{0}{0}$. In the interaction picture with respect to all NV center related terms and the control term, the Hamiltonian becomes
\begin{align}
	H' = \sum_j \omega_j \, \hat{\omega}_j \cdot \vec{I}_j + \frac{m_s}{2} F(t) \sigma_z \sum_j \vec{A}_j \cdot \vec{I}_j,
\end{align}
with the effective Larmor frequency $\omega_j$ and its corresponding rotation axis $\hat{\omega}_j$ originating from the vector $\vec{\omega}_j \defeq \gamma_j B \hat z - m_s \vec{A}_j / 2$, $\hat{\omega}_j\defeq \vec{\omega}_j / \omega_j \defeq \vec{\omega}_j / \abs{\vec{\omega}_j}$. The so-called modulation function $F(t)$ models the action of the control field Hamiltonian $H_c$. It takes the initial value $F(0) = 1$ and alternates the sign at points in time where microwave $\pi$-pulses are applied. Changing again to an interaction picture with respect to the free evolution terms of the nuclear spins yields
\begin{align}
	\label{eq:sys_hamiltonian_interaction_picture}
	H'' &= \frac{m_s}{2} F(t) \sigma_z \sum_j g_j \hat{m}_j(t) \cdot \vec{I}_j + c_j I_j^z,
\end{align}
with the definitions $\hat{m}_j(t) = \cos(\omega_j t) \; \hat{x}_j + \sin(\omega_j t) \; \hat{y}_j$ and $c_j = \vec{A}_j \cdot \hat{\omega}_j$. Note, that we describe the evolution in a new orthonormal coordinate system $\hat{x}_j = \vec{x}_j / g_j$, $\hat{y}_j = \vec{y}_j / g_j, \hat{z}_j = \hat{\omega}_j$ with the vectors $\vec{x}_j = \vec{A}_j - (\vec{A}_j \cdot \hat{\omega}_j) \hat{\omega}_j$, $\vec{y}_j = \hat{\omega}_j \times \vec{A}_j, g_j = \abs{\vec{x_j}} = \abs{\vec{y_j}}$ obtained by Rodrigues' rotation formula, and $I_j^\lambda = \hat{\lambda}_j \cdot \vec{I}_j$ for $\lambda = x,y,z$. When we consider periodic pulse sequences with period $\tau$, the modulation function is also $\tau$-periodic and can thus be decomposed into a Fourier series $F(t) = \sum_{k=0}^\infty a_k \cos(\omega_\seq k t) + b_k \sin(\omega_\seq kt)$ with $\omega_\seq = 2 \pi / \tau$. For the sake of clarity, let us restrict this calculation to the case of even modulation functions $F(t) = \sum_{k=0}^\infty f_k \cos(\omega_\seq k t)$. In that case, and for resonant interaction with one of the nuclear spins, say spin $n$ with $\omega_\seq k_\seq = \omega_n$,  we obtain for the following effective Hamiltonian after eliminating all fast rotating terms
\begin{align}
	H_\text{eff} &= \frac{m_s}{4} f_{k_ \seq} g_n \sigma_z \, I_n^x \label{eq:effHaxy_target} \\ 
	&+ \frac{m_s}{4} f_{k_\seq} \sigma_z \sum_{j \neq n} g_j \left(\cos(\DeltaJ t) I_j^x + \sin(\DeltaJ t) I_j^y \right)  \nonumber \\
	&+ \frac{m_s}{4} \sigma_z \sum_{k \neq k_\seq} \! \! \sum_{j} f_k g_j \left(\cos(\smash{\eta_{jn}^{(k)}t}) I_j^x +\sin(\smash{\eta_{jn}^{(k)}t}) I_j^y \right), \nonumber
\end{align}
with $\DeltaJ = \omega_j - \omega_n$ and $\eta_{jn}^{(k)} = \omega_j - k \omega_n / k_\seq$. The resulting effective Hamiltonian leads to an operation $\zx_{j}(\theta) = e^{-i \theta \sigma_z \otimes I_j^x}$ (Fig.~\ref{fig:fig1}A) if the last two lines of the effective Hamiltonian vanish, which is the case if the system fulfills the following relations (assuming $\omega_j \sim \abs{\gamma_j B} \gg \abs*{\vec{A}_j}$) given by RWA:
\begin{align}
	\abs{\omega_j - \omega_n} &\gg \abs{f_{k_\seq} g_j}, \label{eq:approxCond1} \\
	\abs{\gamma_j B} & \gg \abs{k_\seq g_j}. \label{eq:approxCond2}
\end{align}
These expressions define approximation conditions that limit the usable parameter range with regard to the effective coupling strength $\propto f_{k_\seq}$ by Eq.~\eqref{eq:approxCond1} and the external field $B$ by Eq.~\eqref{eq:approxCond2}. Analogously, a gate of the form $\zy_{j}(\theta) = e^{-i \theta \sigma_z \otimes I_j^y}$ is created by assuming an entirely odd modulation function $F(t) = \sum_{k=0}^\infty f_k \sin(\omega_\seq kt)$. The benefit of using an adaptive sequence for the creation of a gate $\mathcal{A}_j^{x/y} (\theta)$ with $\theta = m_s f_{k_\seq} g_n N \tau / 4$ is the tunability of the Fourier coefficient $f_{k_\seq}$, which extends the set of feasible $\theta$ to a continuity despite the discrete character of the evolution time $N \tau$ of the sequence. The coefficients $f_k$ are determined by varying the pulse positions within the composite pulses (see Appendix \ref{sec:pulsePositions}). Solution for $f_{1_\seq} \in \tfrac{1}{\pi}(-8 \cos(\pi / 9) + 4, 8 \cos(\pi / 9) - 4)$ under the secondary condition $f_k = 0, \; i = 0, 2, 3, 4$ exist, but for values towards the ends of this interval some pulse distances approach zero. Therefore, we are also limited by the microwave Rabi frequency $\Omega_\text{MW} / \tau \ll 1$ to still fulfill the quasi-instantaneous pulse approximation.

We simulate the effective gate using Eq.~\eqref{eq:general_sys_hamiltonian} with additional errors on the microwave control, namely a detuning $\Delta_\text{MW}$ and a multiplicative Rabi frequency error $R_\text{rfe}$ s.t.~$\Omega_\text{MW} \rightarrow \Omega_\text{MW} (1 + R_\text{rfe})$, so that the control Hamiltonian becomes $H_c(t) = m_s \Delta_\text{MW} / 2  + \Omega_\text{MW}(t) (1 + R_\text{rfe})  \left[ \cos(\phi) S_x + \sin(\phi) S_y \right]$. The detuning is assumed to be static, which is a valid assumption if the noise-induced fluctuations of the control parameters are much slower than $\tau$. 
As a figure of merit for the performance of a gate $U$ relative to an ideal gate $U_\text{id}$, we use the fidelity $\fidelity (U, U_\text{id}^{\phantom{\dagger}}) = \abs*{\inlineTr (U_\text{id}^\dagger U)} / [\inlineTr (U_\text{id}^\dagger U_\text{id}^{\phantom{\dagger}}) \tr(U^\dagger U^{\phantom{\dagger}} \! \!)]^{1/2}$. Fig.~\ref{fig:fig1}C shows the simulated fidelity of a $\zx_1$ gate for a nuclear spin pair with orthogonal couplings $A_{\perp,1} = (2 \pi)\, \SI{45.8}{\kilo \hertz}, A_{\perp,2} =  (2 \pi)\, \SI{35.3}{\kilo \hertz}$ and parallel components $A_{\parallel, 1} =   (2 \pi)\, \SI{93.5}{\kilo \hertz}$, \, $A_{\parallel, 2} =   (2 \pi)\, \SI{49.5}{\kilo \hertz}$ (which have been extracted from experimental data \cite{unden2019}) at a $z$-oriented magnetic field of $B = \SI{600}{\gauss}$ and different AXY-8 repetitions $N$ (the AXY-8 sequence has a X-Y-X-Y-Y-X-Y-X structure of composite pulses and therefore a length of $4 \tau$ per repetition). Errors are assumed to be $\Delta_\text{MW} = (2 \pi)\, \SI{350}{\hertz}$, which amounts to temperature fluctuation of $\sim \!\! 5\si{\milli \kelvin}$ (typical temperature stability for cryostats operating at \SI{4}{\kelvin}) with the temperature dependence of ZFS splitting $\dv*{D}{T} = - (2 \pi)\,\SI{74.2}{\kilo \hertz \per \kelvin}$ \cite{acosta2010} and $R_\text{rfe} = 0.25\%$. Assuming a constant detuning represents a worst case scenario, since we approximate the fluctuating process by its maximum $\abs{\Delta (t)} \leq \Delta$.
Fig.~\ref{fig:fig1}C shows that the AXY-based gates can reach infidelities lower than $10^{-3}$ and thus are ideal for serving as building blocks in larger circuits, such as the quantum error correction repetition code as presented in section~\ref{sec:repetitionCode}.

\section{Optimal effective coupling}
\label{sec:relaxingApprox}
In the pursuit of an optimal gate for quantum computing, it is essential to navigate the trade-off between execution speed and fidelity. In practice, this entails selecting an acceptable minimum fidelity value, e.g.~a threshold value at which quantum error correction algorithms can still be reliably executed. One would then attempt to construct gates that achieve this fidelity in the shortest possible execution time. However, the search for the fastest gate in experimental reality is a time-consuming process, as experimental gate characterizations have to resort to lengthy gate tomography procedures. It is therefore necessary to make an analytical preselection of the ideal sequence parameters. The aim of this section is to provide an analytical tool for finding an optimal gate in the high magnetic field regime.
We start from the effective Hamiltonian in Eq.~\eqref{eq:effHaxy_target} and impose the high-field approximation $\abs{\gamma_j B} \gg \abs{k_\seq g_j}$ (Eq.~\eqref{eq:approxCond2}) without the coupling limit condition of Eq.~\eqref{eq:approxCond1}, so that the effective Hamiltonian takes the form
\begin{align}
	\label{eq:hamiltonian_high_field}
	H(t) &=  \sigma_z \otimes \sum_{j} \tilde{g}_j \left[\cos(\DeltaJ t) I_j^x + \sin(\DeltaJ t) I_j^y \right] \nonumber \\ 
	&\eqdef \sigma_z \otimes \sum_{j} H_j(t),
\end{align}
with the spin dependent detuning $\DeltaJ = \omega_j - \omega_\seq$ and $\tilde{g}_j = m_s f_{k_\seq} g_j / 4$. Due to the all-to-one connectivity in this approximated system, the evolution operator obtains a particular structure in the eigenbasis of the $\sigma_z = \dyad{0}{0} - \dyad{1}{1}$ operator (note, that here the labels of $\ket{0}$ or $\ket{1}$ does not represent the respective $m_s$ quantum number):
\begin{align}
	U(t) = &\dyad{0}{0} \otimes \TOop e^{-i \sum_j \int_0^t H_j(s) \, \dd{s}} \nonumber \\ + &\dyad{1}{1} \otimes \TOop e^{i \sum_j \int_0^t H_j(s) \, \dd{s}}
\end{align}
As $[ H_j(t), H_k (t')] = 0 \; \forall t,t' \in \mathbb{R} \; \forall j \neq k$, the exponentials are just a (tensor) product of single spin evolutions and are we can treat each nuclear spin evolution individually. Moreover, In the rotating frame with respect to $U_\Delta(t) = e^{- \smash{\sum_j \DeltaJ I_j^z}}$ each of the local Hamiltonians $H_j(t)$ transform into a constant operator $H_j' = r \, \hat{r} \cdot \vec{I}_j \defeq \tilde{g}_j I_j^x - \DeltaJ I_j^z$. The inverse rotating frame $U_\Delta^{j \dagger}$ brings us back to the original frame, so that the local evolution in the $\dyad{0}{0}$-manifold of the $j$-th nuclear spin is determined by $U_j^{[0]}(t) \defeq e^{i \DeltaJ t I_j^z} e^{-i t r \, \hat{r} \cdot \vec{I}_j}$. Similarly, for the second term corresponding to $\dyad{1}{1}$, the nuclear spin evolution $U_j^{[1]}(t)$ takes the same form, but due to the opposite sign, the vector $\vec{r}$ must be replaced by $\vec{s}$, which is a reflection of $\vec{r}$ with respect to the $yz$-plane $\vec{s} = (-r_x, 0, r_z)^T$, i.e.~$U_j^{[1]}(t) \defeq e^{i \DeltaJ t I_j^z} e^{-i t s \, \hat{s} \cdot \vec{I}_j}$. In summary, we get
\begin{align}
	U(t) &= \dyad{0}{0} \otimes \bigotimes_j e^{i \DeltaJ t I_j^z} e^{-i t r \, \hat{r} \cdot \vec{I}_j} \nonumber \\ &+ \dyad{1}{1} \otimes \bigotimes_j e^{i \DeltaJ t I_j^z} e^{-i t s \, \hat{s} \cdot \vec{I}_j}.
\end{align}
To compare the action of $U(t)$ to our desired reference gate on a target spin $n$ of the form $U_\text{ref}(\theta) = e^{-i \theta \sigma_z \otimes I_j^x}$, that also has a structure 
\begin{align}
U_\text{ref}(\theta) = \dyad{0}{0} \otimes e^{-i \theta I_n^x} + \dyad{1}{1} \otimes e^{i \theta I_n^x},
\end{align}

we define a figure of merit that we will call decoupling efficiency $\deceff_j(t) \defeq \fidelity(\smash{U_j^{[0]} (t)}, \idop)  = \abs*{\tr(\smash{U_j^{[0]} (t)})} / 2$. This function is equal to unity when the operation leaves the $j$-th nuclei untouched - and therefore it represents a local operation $U_j = \idop$; on the other hand it takes values {$\mathcal{D}_j < 1$} when a perturbation on the $j$-th spin occurred. 

An explicit evaluation of the expression $\abs*{\tr(\smash{U_j^{[0]} (t)})} / 2$ yields the formula
\begin{align}
	\label{eq:decoupling_eff}
	\mathcal{D}_j (t) = \abs{\cos(\xi_1) \cos(\xi_2) + \frac{r_z}{\smash{\abs{\vec{r}\,}}} \sin(\xi_1) \sin(\xi_2)}
\end{align}
with $\xi_1 = \DeltaJ t / 2$ and $\xi_2 = \abs{\vec{r}\, } t / 2$. As this formula only depends on the length of the vector $\vec{r}$ and its $z$-component $r_z$, the value is invariant under the exchange $\vec{r} \leftrightarrow \vec{s}$ and we can immediately identify $\tr(\smash{U_j^{[0]} (t)}) = \tr(\smash{U_j^{[1]} (t)})$.

In the case where we want to evaluate the fidelity between a reference gate $U_\text{ref} (\theta)$ and an AXY sequence on resonance the target spin $n$ (i.e.~$k_\seq \omega_\seq$ = $\omega_n$) with duration $T = \theta / \tilde{g}_n$, the trace property $\tr(\bigotimes_k A_k) = \prod_k \tr(A_k)$ allows us to express the achieved fidelity value in terms of a product of decoupling efficiency functions $\mathcal{D}_j \; \forall j \neq n$ ($d$ is the number of nuclei):
\begin{widetext}
\begin{align}
	\fidelity(U_\text{ref}(\theta), U(T)) &=  \frac{1}{2^{1+d}} \Bigl \lvert \tr \Big( \sum_{k=0,1} \dyad{k}{k} \otimes \! \! \! \bigotimes_{1 \leq j < n} \! \! U_j^{[k]}(T) \otimes \underbrace{e^{(-1)^{1+k} i\theta I_n^x} U_n^{[k]}(T)}_{= \idop} \otimes \bigotimes_{j > n} U_j^{[k]}(T) \Big) \Bigr \rvert \nonumber \\
	 &= \frac{1}{2} \Big( \prod_{j \neq n} \tfrac{1}{2} \abs*{\tr(\smash{U_j^{[0]} (T)})} + \prod_{j \neq n} \tfrac{1}{2} \abs*{\tr(\smash{U_j^{[1]} (T)})} \Big) = \prod_{j \neq n} \deceff_j (T) \label{eq:fidelity_as_D_product}
\end{align}
\end{widetext}
However, the description in terms of decoupling efficiency does not suffice yet for the use in our suggested task of selecting the fastest gates as we haven't yet incorporated the adaptive character of the sequence into the discussion. Assume that a gate $U_\text{ref}(\theta)$ with rotation angle $\theta$ is implemented during an evolution time $t = 4 \tau N$ (in the case of AXY-8). As the effective rotation angle on an AXY-8 sequence is given by $\phi_\text{AXY-8} = m_s f_{k_\seq} g_n t / 4$ (Eq.~\eqref{eq:effHaxy_target}), we need to adjust the Fourier coefficient $f_{k_\seq}$ inversely proportional with time $f_{k_\seq} \propto t^{-1}$ when we want to evaluate the fidelity for the same rotation angle $\theta$ but variable filter function shapes, i.e.~for different evolution times $t$. Consequently, the effective coupling constants must obey $\tilde{g}_j \propto t^{-1}$ as well. Plugging this into Eq.~\eqref{eq:decoupling_eff} with $r_x \! = \! \tilde{g}_j \cramped{( \tilde{g}_j^2 + \DeltaJ^2 )^{-1/2}}$ and $r_z \! = \! \cramped{- \DeltaJ ( \tilde{g}_j^2 + \DeltaJ^2 )^{-1/2}}$ as well as rescaling to a dimensionless time $\vartheta = \DeltaJ  t / 2$ we finally get the desired decoupling efficiency functions
\begin{align}
	\label{eq:decoupling_efficiency_adaptive_sequence}
	\deceff_j(\vartheta)\! = \! \left| \cos(\vartheta) \cos(\mu (\vartheta)) \! + \! \frac{\vartheta}{\mu (\vartheta)} \sin(\vartheta) \sin(\mu (\vartheta)) \right|.
\end{align}
with $\mu (\vartheta) = \cramped{({\vartheta}^2 + (g_j \phi / 2 g_n )^2)^{1/2}} \; \forall j \neq n$ .
The relation of gate fidelity and decoupling time $T = 4 \tau N$ is plotted in Fig.~\ref{fig:fig1}D with an exemplary system consisting of (2+1) spins and the simulated fidelity values and the analytically calculated fidelity (Eqs.~\eqref{eq:fidelity_as_D_product} and \eqref{eq:decoupling_efficiency_adaptive_sequence}). We can observe high-fidelity solutions for repetition numbers that would have been excluded by the approximation condition in Eq.~\eqref{eq:approxCond1}, so it is beneficial to study the decoupling efficiency functions for gate-time-optimized sequence parameters. A simulation with and without inter-nuclear coupling shows that it has a negligible effect on the result for the timescale of the gate. The discrepancy for higher repetition numbers is therefore originated from the third term in Eq.~\eqref{eq:effHaxy_target}, so the higher harmonics cause a slight deterioration of the gate for longer evolution times due to insufficient fulfillment of the high field condition \eqref{eq:approxCond2}.
\section{Repetition code for quantum error correction}
\label{sec:repetitionCode}
\begin{figure}[htb]
	\centering
	\includegraphics[width=\linewidth]{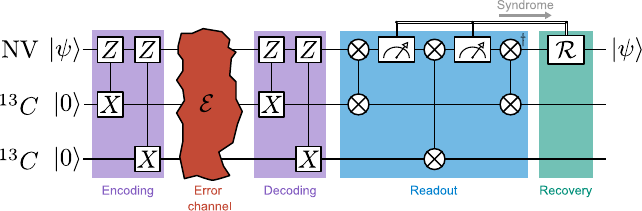}
	\caption{Phase error repetition code for a 2+1-qubit system. The $\iswap$ gates in the readout part of the protocol are represented by the $\bigotimes$-notation and its inverse is marked with an additional dagger symbol.}
	\label{fig:repetitionCode_circuit}
\end{figure}
Despite all efforts to develop methods for generating high-fidelity gates, these operations will always have small deviations from the idealized gate because it is impossible to account for all sources of crosstalk, decoherence, and relaxation in the model. To further reduce the logical qubit errors, it is inevitable to complement the gate optimization efforts with quantum error correction.
Here we construct a basic example of a quantum error correction code and show that such a code can be constructed with the designed electronic-nuclear two-qubit gates. Specifically, we discuss a version of the repetition code for an NV center equipped with an auxiliary system of two $\cxiii$ nuclear spins - a (2+1)-qubit system - to correct phase errors. As a toy-model, consider a single qubit interacting with a not further specified environment via a dephasing interaction $H_\text{int} \propto \sum_{j \geq 0} {\sigma_z}^{(j)} \otimes B$. If the  Born-Markov approximation is applicable, we can formulate a Lindblad master equation with the dephasing dissipator \cite{breuer2010}
\begin{align}
	\mathfrak{D}(\rho) = \gamma (\sigma_z \rho \sigma_z -\rho),
\end{align}
where $\gamma$ the dissipation rate. This dissipator leads to a quantum channel of the form {$\Lambda_j (\rho) = (1-p) \rho + p \,  \cramped{{\sigma_z}^{(j)}} \rho {\sigma_z}^{(j)}$} with probability $p = (1 - e^{-\gamma t}) / 2$. For simplicity we assume equal error probabilities on all qubits and we also view the dephasing of the spins as independent, so that the final error channel is given by
\begin{align}
	\label{eq:recoveryCode_errorChannel}
	\mathcal{E}(\rho) = \prod_{j \geq 0} \Lambda_j (\rho)  = \prod_{j \geq 0} \left[ (1-p) \rho + p \sigma_z^{(j)} \rho \sigma_z^{(j)} \right],
\end{align}
where index $j = 0$ represents an operation on the electron spin. Consequently, the probability for a single qubit error is then given by $3 p (1-p)^2$. 

The presented protocol can be also rewritten to correct flip errors $\sigma_x^{(j)}$ by adding local operations on the NV center. Encoding and decoding in the standard form of the repetition code is done with CNOT gates \cite{nielsen2010} and it is in general possible to compose a CNOT gate with $\zx_j(\theta)$ or $\zy_j(\theta)$ operations. However, we refrain from this approach as it would involve local rotations on the nuclear spins which are temporally extensive for spectrally dense spin ensembles. A more suitable encoding for our gate set is $U_\text{enc} = \zx_2(\pi / 2) \zx_1(\pi / 2)$. It requires only $\zx$-gates (Fig.~\ref{fig:repetitionCode_circuit}), which can be performed with high-fidelity, and provides the ability to detect and reverse phase errors as it can be seen from the syndrome table in Tab.~\ref{tab:recoveryCode_syndrome}.
\begin{table}[htb]
	\def\ls{1pt}
	\centering
	\begin{tabular}{ !{\vrule width \ls} c !{\vrule width \ls} c|c|c !{\vrule width \ls}}
		\noalign{\hrule height \ls}
		\textbf{Error} $\mathbf{E_k}$ & $\mathbf{U_\text{enc}^\dagger E_k U_\text{enc}}$ & \textbf{Syndrome} & \textbf{Recovery} $\mathbf{\mathcal{R}_k}$ \\ 
		\noalign{\hrule height \ls}
		$\mathbf{\idop}$     & $\idop$                                                   &  (0,0)   & $\idop$          \\ \hline
		$\mathbf{Z_0}$       & $\sigma_x^{(0)} \sigma_x^{(1)} \sigma_x^{(2)}$            &  (1,1)   & $\sigma_x^{(0)}$ \\ \hline
		$\mathbf{Z_1}$       & $\sigma_z^{(0)} \sigma_y^{(1)}$                           &  (1,0)   & $\sigma_z^{(0)}$ \\ \hline
		$\mathbf{Z_2}$       & $\sigma_z^{(0)} \sigma_y^{(2)}$                           &  (1,1)   &$\sigma_z^{(0)}$  \\ 
		\noalign{\hrule height \ls}
	\end{tabular}
	\caption{Syndrome table for the 2+1-qubit recovery code for the encoding $U_\text{enc} = \zx_2(\pi / 2) \zx_1(\pi / 2)$. It can be seen that every error originating from the channel in \eqref{eq:recoveryCode_errorChannel} is mapped to distinguishable orthogonal subspaces. For every error $E_k$ the operation after decoding $U_\text{enc}^\dagger E_k U_\text{enc}$ is given together with the corresponding syndrome for an input state $\ket{\psi} \otimes \ket{0} \otimes \ket{0}$ and the suitable recovery operation $\mathcal{R}_k$ to retrieve the input information $\ket{\psi}$.}
	\label{tab:recoveryCode_syndrome}
\end{table}
Acquiring the syndrome, i.e.~readout of the nuclear spin states, can also be done by employing the AXY-gates to create swap operations. To be specific, as the realization of $\swap$-gates also requires local operations on nuclear spins, we restrict ourselves to the $\iswap$-gate, which can be realized by electron spin rotations $X_{\pi / 2}, Y_{\pi / 2}$ and two-qubit gates only, because $\iswap = e^{i \frac{\pi}{4} (\sigma_x \sigma_x + \sigma_y \sigma_y)}$ between the electron spin and the $j$-th nuclear spin can be rewritten as:
\begin{align}
	\iswap_{j}= Y_{\frac{\pi}{2}} \zx_{j} \left(\tfrac{\pi}{2}\right) Y_{\frac{\pi}{2}}^\dagger X_{\frac{\pi}{2}}^\dagger \zy_{j} \left(\tfrac{\pi}{2}\right) X_{\frac{\pi}{2}}.
	\label{eq:iswap_identity}
\end{align}
The conditional phases imposed by the $\iswap$ operation are irrelevant for the storage of the quantum information $\ket{\psi}$ as the gate inversion can be done by $\iswap^\dagger$ as depicted in Fig.~\ref{fig:repetitionCode_circuit}. The measurements are assumed to be ideal in the sense that they are noiseless and return outcomes with probabilities according to a statistic following Born's rule.

The corruption of the recovered quantum information $\ket{\psi}$ has two causes.  Firstly, erroneous operations during encoding, decoding and the readout process lead to incorrect probability amplitudes of  $\ket{\psi}$.  In addition, the faulty two-qubit gates lead to further faulty terms in the Pauli decomposition of the operators in the quantum circuit, so that these can influence the probability amplitudes of the measurements and thus return incorrect syndromes, which consequently lead to invalid correction operations.

\begin{figure}[htb]
	\centering
	\includegraphics[width=\linewidth]{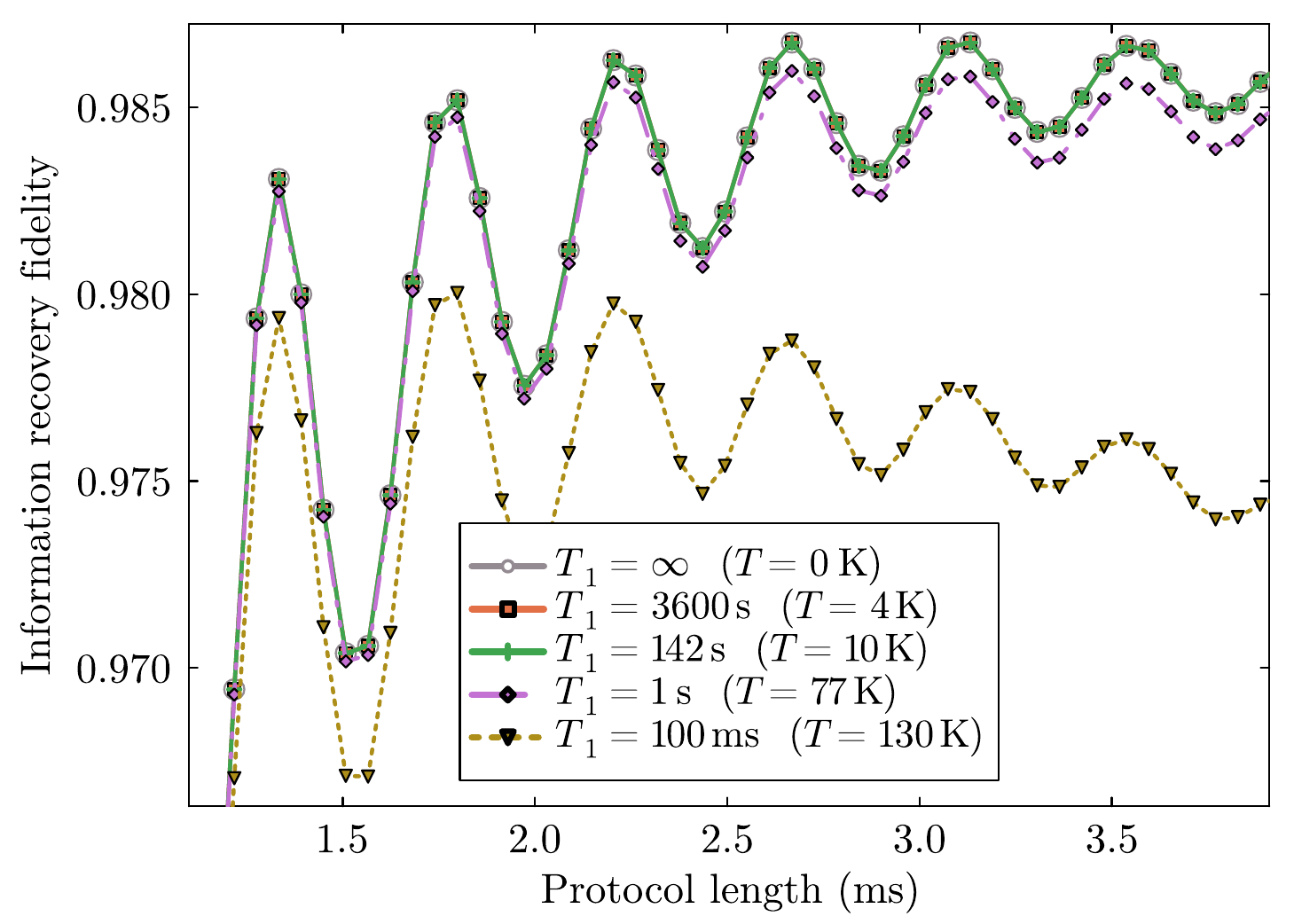}
	\caption{Information retrieval fidelity for the phase error repetition code based on the AXY-gate encoding. The longitudinal relaxation of the electron spin was simulated according to Eq.~\eqref{eq:dissipator_t1} with corresponding temperature $T$. Relaxation rates and temperature data has been extracted from experimental data. Even for lower repetition numbers, and hence lower protocol execution time, the code achieves a state fidelity of \SI{>0.985}{\percent} in the low temperature regime. The variation of  the protocol length are realized by increasing or decreasing the repetition numbers for the AXY-gates.}
	\label{fig:repetitionCode_fidelity}
\end{figure}

We simulate the success rate of this protocol by determining the average state fidelity between the input and the output state of this protocol, i.e.%
\begin{align}
	\overline{\mathcal{F}}_\text{corr} (t) = \int_\psi \dd{\psi} \inlineTr(\dyad{\psi}{\psi} \Gamma[\dyad{\psi}{\psi}]),
\end{align} 
where $\Gamma [\rho]$ is the non-unitary quantum channel of the protocol and $\dd{\psi}$ is the Haar measure with normalization $\int_\psi \dd{\psi} = 1$ \cite{nielsen2002a}. Fig.~\ref{fig:repetitionCode_fidelity} shows that with this protocol and the speed-optimized gates presented in Section~\ref{sec:relaxingApprox} allow an average correction fidelity above \SI{98.5}{\percent} for an error probability p=\SI{5}{\percent} in a decoherence free system ($T_1 = \infty$).
As already stated in  Section~\ref{sec:gates_basics}, the execution speed of the AXY gates, fundamentally determined by the maximally possible value for $f_{k_\seq}$, is upper bounded by the duration of the MW-$\pi$-pulses, and thus by the NV center Rabi frequency $\Omega_\text{MW}$.

Even for the repetition code, which is the smallest code for error correction that one can construct, the total correction sequence length is in the order of milliseconds. To still be able to confidently achieve the high-fidelity regime, the faster decoherence processes due to longitudinal electron spin relaxation should be included into the model. For this we include the photon-induced excitation and decay of the NV center's spin by adding the dissipators
\begin{align}
	\mathfrak{D}(\sigma_-) \! &= \! \lambda (1 \! + \! \expval{n(\omega_\text{NV}, T)}) \! \left[ \sigma_- \rho \sigma_+ \! + \! \frac{1}{2} \! \left\{ \sigma_+ \sigma_-, \rho \right\}\right], \nonumber \\
	\mathfrak{D}(\sigma_+) \! &= \! \lambda  \! \expval{n(\omega_\text{NV}, T)} \! \left[ \sigma_+ \rho \sigma_- \! + \! \frac{1}{2} \! \left\{ \sigma_- \sigma_+, \rho \right\}\right]. \label{eq:dissipator_t1}
\end{align}
The average photon number $\expval{n(\omega_\text{NV}, T)}$ is given by the Bose-Einstein distribution $\expval{n(\omega_\text{NV}, T)} = 1 / [\exp(\hbar \omega_\text{NV} / k_B T) -1]$ and the parameter $\lambda$ is used to fit the model to a desired relaxation rate $T_1$ by
\begin{align}
	1 / T_1^\text{exp} = \lambda  \expval{n(\omega_\text{NV}, T)}.
\end{align}
With this relation, the dissipation condition can be engineered to agree with experimentally measured noise. In Fig.~\ref{fig:repetitionCode_fidelity} we show the simulations for dissipation strengths and temperatures noted by Refs.~\cite{abobeih2018, bar-gill2013, jarmola2012}. The simulation result for this data, also depicted in Fig.~\ref{fig:repetitionCode_fidelity}, shows that an environment specified by a temperatures even up to $\SI{77}{\kelvin}$ generating a electron spin longitudinal relaxation up to $T_1 = \SI{1}{\second}$ does not cause a significant deviation from the decoherence-free setup for the efficiency of the repetition code. 

\section{Conclusion \& Outlook}
We have demonstrated the creation of high-fidelity two-qubit gates that exhibit robustness against control errors, which enabled the implementation of a small error correction code for phase correction. Additionally, with the presented method in this work, we are able to identify the best sequence parameters to tailor the gate to our fidelity- and speed requirements. While this parameter search may be feasible by trial-and-error for a small number of nuclear spins, this method is indispensable when scaling up the system size.
Compared to the shortest theoretically possible gate time $T_\text{min}$ for a two spin system that does not require decoupling, we find gates with $T \approx 2.2 \, T_\text{min}$ to reach an infidelity $< 10^{-2}$ and $T \approx 4.4 \, T_\text{min}$ for infidelities below $10^{-3}$.

Another approach to improve selectivity, the so called soft control method \cite{haase2018}, is introducing a time-dependent coupling constant $f_{k_\seq}(t)$ that follows a Gaussian trajectory (see Appendix \ref{sec:soft_control}). This method could be particularly beneficial in spectrally dense nuclear spin environments, as it reduces the need for drastic coupling strength reductions. Furthermore, this modification may prove more advantageous in instances where inaccuracies in the system parameter characterization occur, as it is more robust against such errors.
However, this approach comes with certain drawbacks. Due to the nature of pulse sequences, that requires generating a Gaussian profile through a piecewise constant function, the robustness of the sequence may be compromised with respect to microwave pulse errors.

Future steps left for future work is the development of a full error correction scheme (e.g., the five-qubit code) or a surface code, where NV centers serve as readout qubits and carbon nuclei form the data qubits.

\section{Acknowledgments}
This work was supported by the German Federal Ministry of Science (BMBF) under the project SPINNING (Grant No.~\texttt{13NIG215}) and CoGeQ (Grant No.~\texttt{13N16101}). Discussions with Jan Haase, Martin Korzeczek, Yu Liu, Jonas Breustedt and Jorge Casanova have greatly enriched the work.

During preparation of this work woe became aware of \cite{finsterhoelzl2024} which applies similar ideas to different control scenarios.

%
%
%
\appendix

\section{Derivation of pulse positions}
\label{sec:pulsePositions}
The modulation function $F(t)$ of the pulse sequence with pulses occurring at times $t_i$ within a period $\tau$ can be described via the Heaviside step function $\Theta$ as
\begin{align}
	\eval{F(t)}_{t \in [0, \tau]} = 1 + 2 \sum_{i=1}^{10} (-1)^i \Theta (t - t_i).
\end{align}
The coefficients of the respective Fourier series $F(t) = \sum_{k=0}^\infty a_k \cos(\omega_\seq k t) + b_k \sin(\omega_\seq k t)$ (with $\omega_\seq = 2 \pi / \tau$) evaluate to
\begin{align}
	a_k &= \frac{2}{\tau} \int_0^\tau F(t) \cos(\omega_\seq k t) \dd{t} \nonumber \\ 
	    &= \frac{2}{\pi k}  \sum_{i=1}^{10} (-1)^{i+1} \sin (\omega_\seq k t_i), \\
	b_k &= \frac{2}{\tau} \int_0^\tau F(t) \sin(\omega_\seq k t) \dd{t} \nonumber \\ 
	    &= \frac{2}{\pi k}  \sum_{i=1}^{10} (-1)^{i} (\cos (\omega_\seq k t_i) - 1) \nonumber \\ 
	    &= \frac{2}{\pi k}  \sum_{i=1}^{10} (-1)^{i} \cos (\omega_\seq k t_i).
\end{align}
We can use the Chebyshev polynomials of the first and second kind $T_n$ and $U_n$, an use the fact that they fulfill the identities
\begin{align}
	\sin(kx) &= U_{k-1}(\cos(x)) \sin(x), \\
	\cos(kx) &= T_{k}(\cos(x)),
\end{align}
to express the Fourier coefficients as a function of the first harmonic only:
\begin{align}
	a_k &= \frac{2}{\pi k}  \sum_{i=1}^{10} (-1)^{i+1} \, U_{k-1}(\cos(2 \pi x_i)) \sin(2 \pi x_i), \\
	b_k &= \frac{2}{\pi k}  \sum_{i=1}^{10} (-1)^{i} \, T_k (\cos (2 \pi x_i)),
\end{align}
with the dimensionless positions $x_i = t_i / \tau \in [0, 1]$. Solving these equations for $\{ x_i \}$ gives the pulse positions for desired coefficients $\{a_k \}$, $\{b_k \}$. For example in the case of a even modulation function and selecting $a_1 \neq 0$ while $a_k = 0$ for $i = 2,3,4$, these equations can be solved analytically as demonstrated in Ref.~\cite{casanova2015}. The reformulation into Chebyshev polynomials is not strictly necessary but may help in solving the equations.

\section{Estimation of coupling strengths}
\begin{figure}[tb]
	\centering
	\includegraphics[width=\linewidth]{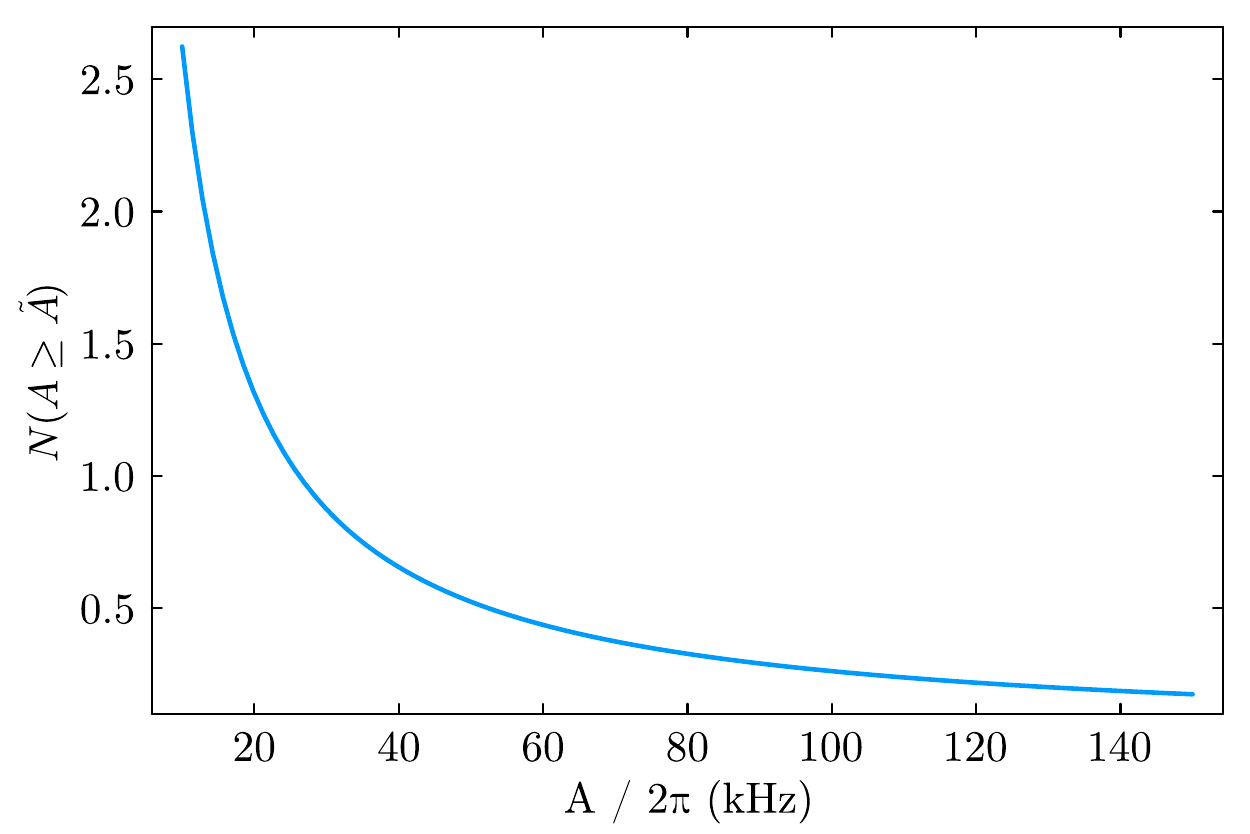}
	\caption{Estimation of coupling strength.}
	\label{fig:couplingStrengthEstimation}
\end{figure}
In this work, experimental data from \cite{unden2019} has been used as values for the coupling strengths in the simulations. In order to determine whether such a configuration in the experiment is just a `lucky find' or a regular occurrence, we try to roughly estimate the number of spins in the surroundings of an NV center that possesses the hyperfine vectors mentioned in the paper. The hyperfine vector given by (in units of $\hbar$)
\begin{align}
	\vec{A}_j = \frac{\mu_0 }{4 \pi} \frac{\hbar \gamma_e \gamma_j }{r^3} \left(\hat{z} - 3 \cos \theta \, \hat{r}\right)
\end{align}
may be written by a scalar component $A(r) = \alpha / r^3$ and a geometric factor $\left(\hat{z} - 3 \cos \theta \, \hat{r}\right)$. We try to roughly estimate the probability of coupling strengths that can occur in a diamond with  $p_{13C} = 1.1 \%$ natural abundance of ${}^{13}C$. We disregard the discrete diamond lattice structure and instead use the density of carbon atoms $\rho = 8 / V_\text{UC}$ with the unit cell volume $V_\text{UC}$. Thus, the total number of $\cxiii$-atoms in a sphere with radius $R$ is given by
\begin{align}
	N(R) = \frac{4}{3} \pi R^3 \rho \, p_{13C}
\end{align}
The spin density with respect to the coupling strength $A$, denoted as $\eta (A)$ is given by
\begin{align}
	\eta (A) \defeq \dv{N(A)}{A} = \dv{N}{r} \dv{r}{A} = \dv{N}{r} \! (r(A)) \, \dv{r(A)}{A},
\end{align}
with the inverse function $r(A) = A^{-1}(r)$. The integration
\begin{align}
	\label{eq:numberN_coupling_estimation}
	N(A \geq \tilde{A}) = \int_{\tilde{A}}^\infty \eta(A) \dd{A} = - \frac{4}{3} \pi \rho p_{13C} \alpha \tilde{A}^{-1}
\end{align}
returns the number of spins that have a coupling greater as $\tilde{A}$. This quantity is plotted in Fig.~\ref{fig:couplingStrengthEstimation} over $\tilde{A}$. For the hyperfine vectors used in the main text, we find $N(A \geq A_1) = 0.48$ and$N(A \geq A_2) = 0.81$, hence there is a probability of roughly $39 \%$ to find two nuclear spins with a coupling strength $\abs*{\vec{A}_j}$ of at least the value of $A_1$ and $A_2$ respectively. Note that this is just a crude estimation as the geometrical factor has been completely neglected in the discussion, This vectorial multiplicative factor has a norm of $\sqrt{1 + \cos^2 \theta} \in [1, \sqrt{2}]$ and thus causes a deviation between the scalar component $A(r)$ and the actual coupling $\abs*{\vec{A}(\vec{r})}$.

\section{Filter functions in the context of gates}
\begin{figure}[ht]
	\centering
	\includegraphics[width=1\linewidth]{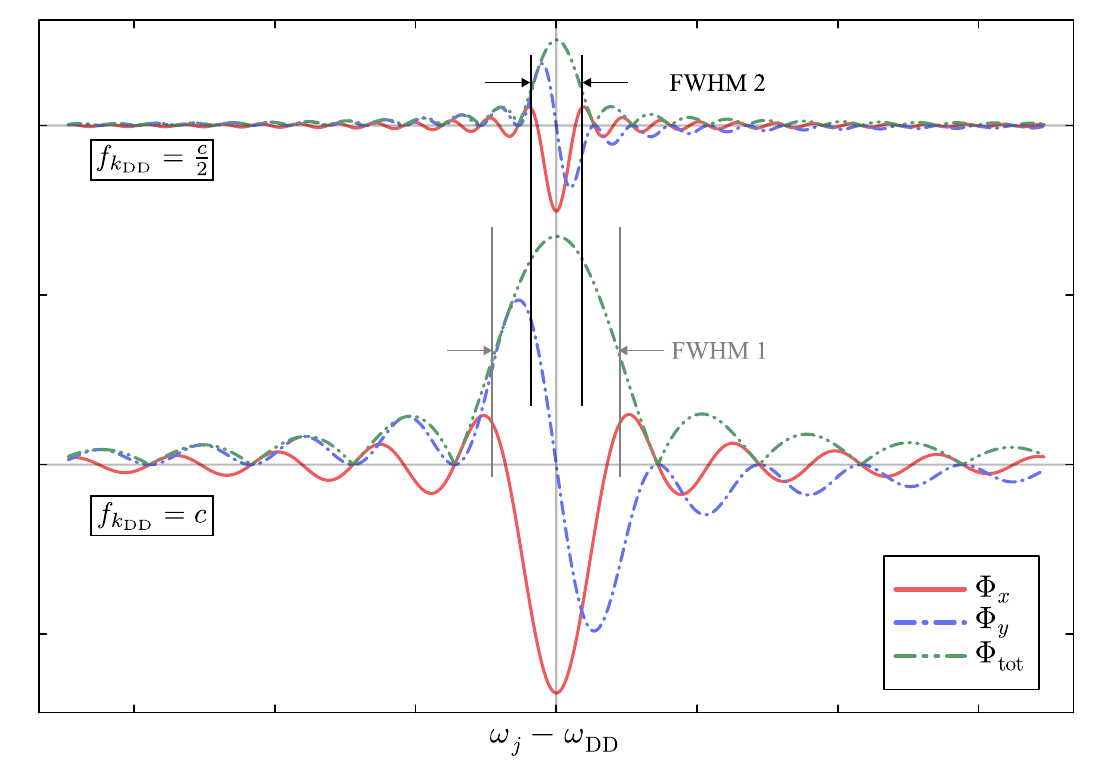}
	\caption{Relation between filter functions and coupling strength. The filter functions of a pulse sequence with period $\tau = 2 \pi / \omega_\seq$ is plotted against the nuclear spin Larmor frequency $\omega_j$ for a given Fourier coefficient $f_{k_\seq}$ (bottom) and a coefficient reduced by a factor of two (top). The reduction of the effective coupling leads to a smaller but sharper peak. }
	\label{fig:filter_function}
\end{figure}
\label{sec:filter_functions}
Filter functions provide a convenient and intuitive concept to investigate the effect of pulse sequences to a system. This method is widespread in the quantum sensing community where filter function give insights into the sensitivity and decoupling capabilities to certain frequency components to incoming noise \cite{biercuk2011, ajoy2011,pham2016a}. This concept is also extendable for analysis in gate design \cite{albrecht2015, albrecht2015a}. Let us consider a single nuclear carbon spin interacting with an NV center and start from the interaction Hamiltonian in Eq.~\eqref{eq:sys_hamiltonian_interaction_picture}
\begin{align}
	H'' &=  F(t) \tilde{g} \left\{ \cos(\omega t) \sigma_z \otimes I_x + \sin(\omega t) \sigma_z \otimes I_y\right\},
\end{align}
where all prefactors have been absorbed into the constant $\tilde{g}$. By evaluating the Magnus expansion in the low coupling limit ($\omega \gg \tilde{g}$) up to the first order, we obtain the evolution operator $	U \simeq \exp(-i \tilde{g} t \sigma_z \otimes \sigma_\varphi)$ with
\begin{align}
\sigma_\varphi = \frac{\Phi_x(P, \tau_\text{DD})}{\Phi_\text{tot}(P, \tau_\text{DD})} I_x + \frac{\Phi_y(P, \tau_\text{DD})}{\Phi_\text{tot}(P, \tau_\text{DD})} I_y.
\end{align}
Here, $P$ is the number of pulses and $\tau_\text{DD}$ is the period of the pulse sequence. The functions $\Phi_x, \Phi_y$ are the filter functions defined by $\Phi_x \defeq \Re(\chi(t))$ and $\Phi_y \defeq \Im(\chi(t))$ through the function
\begin{align}
	\chi (t) = \int_{0}^{t} \dd{s} F(s) e^{i \omega s}.
\end{align}
The total filter $\Phi_\text{tot}$ is just the absolute value of $\chi$, i.e.~$\abs{\chi (t)} = \sqrt{\Phi_x^2 + \Phi_y^2}$. In the context of adaptive dynamical decoupling sequences, a reduction of the effective coupling is enhancing spectral selectivity of the sequence as demonstrated in Fig.~\ref{fig:filter_function}. Although this concept is not used for the analysis in this work, it provides a clear illustration of the conflict of objects between speed and selectivity in the gate design. 

\section{Gate fabrication using soft control}
\label{sec:soft_control}
\begin{figure}[b]
	\centering
	\includegraphics[width=1\linewidth]{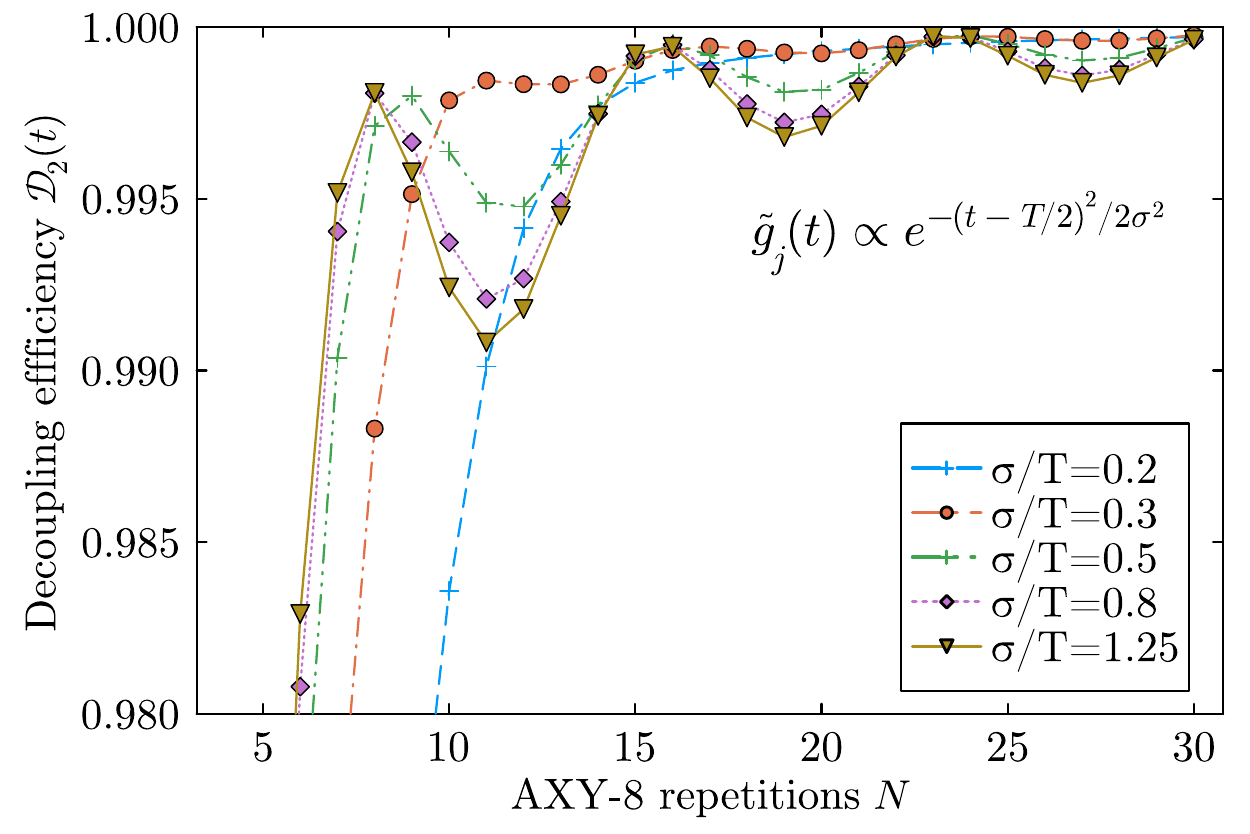}
	\caption{Decoupling efficiency functions $\deceff_2(N)$ for different Gaussian amplitude modulations $\sigma / T$. The amplitude modulation is assumed to be continuous.}
	\label{fig:soft_control}
\end{figure}
An extension to a AXY sequence with $N$ repetitions is the so-called \textit{soft control method} \cite{haase2018}, where the Fourier coefficient $f_{k_\seq}$ is not kept constant throughout the sequence, but is instead modulated with a Gaussian profile. Starting from Eq.~\eqref{eq:hamiltonian_high_field}, we replace $\tilde{g}_j \defeq c_j f_{k_\seq} \rightarrow \tilde{g}_j(t) = c_j f(t)$ and assume a Gaussian profile for f(t), i.e.
\begin{align}
	f(t) = f_0 \exp(-\frac{(t-\frac{T}{2})^2}{2 \sigma^2}),
\end{align}
with $T$ being the length of the sequence. As we intend to perform a rotation on the target spin with index $n$ by some angle $\Theta$, we need to fulfill the relation $\Theta = \int_{0}^{T} \tilde{g}_n(t) \dd{t}$. Thus, for a given $T = 4 \tau N$ and profile variance $\sigma$, we can find the suitable amplitude by
\begin{align}
	f_0 = \frac{\Theta}{\sqrt{2 \pi} \sigma c_n \erf \big(\frac{T}{\sqrt{8} \sigma} \big)}.
\end{align}
Note, that the amplitude of the Gaussian is bounded by $\abs{f_0} < (8 \cos(\pi / 9) - 4) / \pi \approx 1.1$. Using values close to that bound might cause problems in a practical implementation with finite-width pulses, as overlapping of microwave pulses can occur for low Rabi frequencies.

A numerical evaluation of the decoupling efficiency function $\deceff_2(N)$ for various variances $\sigma / T$, shown in Fig.~\ref{fig:soft_control} for the gate $\exp(-i \tfrac{\pi}{2} \sigma_z I_1^x)$, allows us a comparison to Fig.~\ref{fig:fig1}D. We observe that larger variances, i.e. amplitude modulations that don't differ vastly from the constant case, retain the previously observed oscillations in the short-sequence regime. However, in cases where most of the Gaussian profile is incorporated in the modulation, we see a suppression of this features, which is a witness of improved decoupling. 
However, one has to emphasize that we assumed a continuous modulation of the coefficient $f$ which is not possible for pulse sequences, where the modulation function is restricted to piecewise constant maps. Obviously, the approximation of a function $f(t)$ by a piecewise constant function $\tilde{f}(t) = c_i\,  \forall t \in [t_i, t_i + \Delta t)$ gets more accurate for a decreasing sampling interval length $\Delta t \rightarrow 0$. In case of a AXY-8 sequence, there are only a few reasonable choices for $\Delta t$, namely $4 \tau$ (a full AXY-8 repetition), $2 \tau$ (a AXY-4 block), $\tau$ (a composite XY pulse) and $\tau / 2$ (a single composite pulse, either X or Y). The less X or Y pulses we choose in our sampling interval, the more robustness features of the sequence we loose \cite{casanova2015}. Hence, the idealized calculation in Fig.~\ref{fig:soft_control} might loose its validity especially for very short sequence lengths $T$. 

\bibliography{literature.bib}
\end{document}